# Superconductivity at high $T_c$ in neodymium-doped 1111-SrFeAsF system

S.V. Chong,[a,*] T. Goya,[b] H. Yamaguchi,[b] K. Kadowaki[a,b]

[a]*Institute for Materials Science, University of Tsukuba, 1-1-1, Tennodai, Tsukuba, Ibaraki 305-8573, Japan*

[b]*Graduate School of Pure and Applied Sciences, University of Tsukuba, 1-1-1, Tennodai, Tsukuba, Ibaraki 305-8573, Japan*

**Abstract**

Polycrystalline $Sr_{1-x}Nd_xFeAsF$ samples were prepared at various Nd-doping levels using both a stoichiometric mixture of the starting materials and in slight excess amounts of FeAs. Susceptibility and resistivity of the samples were studied down to 4 K revealing a probable coexistence of superconductivity and a magnetic ordering. Temperature dependence of resistivity for all the Nd-doped samples shows the presence of a transition below 15 K most likely originating from the magnetic ordering of Nd moments, while the spin-density-wave anomaly at 175 K survives up to 0.35 Nd-doping. Superconductivity only occurs above 0.40 Nd-doping with onset maximum $T_c$ reaching as high as 52 K.

The discovery of superconductivity in iron-based compounds early last year has generated a huge amount of research efforts in this field [1]. To date, more than five types of iron-based superconductors have been revealed including one being isostructural with the initially discovered iron oxypnictides (*Ln*FeAsO, where *Ln* = rare-earths), but with the oxygen constituent fully replaced by fluorine and the *Ln* sites replaced by divalent cations. As in the 122-FeAs, this so-called 1111-fluoroarsenide system comes in the form of SrFeAsF, CaFeAsF, BaFeAsF and EuFeAsF [2-4]. Superconductivity can be induced upon the replacement of some of the divalent cations with trivalent rare-earth cations (La, Pr, Nd, and Sm) [4-7] or by substituting some of the Fe with Co in the FeAs layers [3,8]. The maximum $T_c$s' achieved in these compounds are only slightly higher than those of the corresponding rare-earth oxypnictides with the highest reported at 57.4 K in Nd-doped CaFeAsF [6]. In this paper, we report on the realization of superconductivity in Nd-doped SrFeAsF above 51 K. We also report on a unique technique to control the concentration of Nd in this quaternary system by adding fixed excess amounts of FeAs which acts like a solvent in assisting the doping of Nd into the $[SrF]^+$ layers.

Polycrystalline samples with nominal composition of $Sr_{1-x}Nd_xFeAsF$ were prepared by solid state reaction using $SrF_2$, SrAs, NdAs, As, and Fe as starting materials. Two methods were used to prepare the $Sr_{1-x}Nd_xFeAsF$ samples – type I uses a stoichiometric mixture of the starting materials, and type II uses fixed amounts of extra added FeAs according to the formula $(Sr_{1-x}Nd_xF)_{20-2y}(FeAs)_{20+y}$, where y = 2.5 and 5 were used. In brief, all samples were prepared by first heating the thoroughly ground mixed powders of the starting materials placed in sealed quartz tubes containing partially reduced argon gas at a rate of 50°C/h to 1000 °C for 24 h. The sintered products were then ground, pressed into pellets and heated at 1000 °C for 20 to 30 h also under reduced argon gas atmosphere; this step was repeated at least 5 times using type I method, and at least 3 times using type II. The crystal structure of the resulting samples was characterized by powder x-ray diffraction (XRD) using Cu Kα radiation. XRD indicates all the samples contain $SrF_2$, NdAs, and smaller amounts of FeAs and $Fe_2As$ impurity phases, which in total accounts for no more than 40% volume fraction. The concentration of Nd incorporated into the quaternary was determined by energy-dispersive x-ray spectroscopy (EDS), which also confirms the impurities detected in XRD.

Fig. 1 shows the temperature dependent resistivity (*R-T*) of $Sr_{1-x}Nd_xFeAsF$ measured using a standard four-probe dc method. The displayed Nd compositions are averaged values obtained from EDS. $Nd_x$ = 0.23 and 0.50 samples were prepared via type I method. Samples with $x$ = 0.36 and 0.43 were prepared from initial Nd nominal composition of 0.4 and 0.5, respectively, and with y = 2.5. Sample $Nd_x$ = 0.48 was synthesized from starting composition $x$ = 0.5 and y = 5. The resistivity trace for $x$ = 0.23 indicates a steep drop in resistivity at 175 K as the sample was cooled down. This is attributed to the magneto-structural transition of the spin-density wave (SDW) instability which is observable in all non-doped SrFeAsF (not shown here) samples [2,5,7]. At the lower temperature region around 9 K another transition is observed, which is most likely originating from the magnetic ordering of Nd moment also observable in all our temperature dependent magnetization measurements (*M-T*). Doping $Nd_x$ = 0.36 into SrFeAsF is sufficient to completely suppress the SDW anomaly at 175 K but no evidence of superconductivity was observed, which might be caused by the pronounced low temperature transition as the resistivity rises sharply peaking at around 11 K before decreasing down to 4 K (left inset in Fig. 1). Upon $x$ = 0.43 Nd-doping, a sharp drop in resistivity occurs evident of a superconducting transition with an onset $T_c$ commencing just above 43 K. At higher Nd-doping ($x \geq 0.46$), even higher onset $T_c$s' are realized reaching to a maximum value just below 52 K. It is noticed that the low temperature transitions are still observable in the superconducting state but shift to a higher temperature position at 13.4 K and the transitions are less pronounce compared with the non superconducting state. All the superconducting transitions are supported by *M-T* measurements where both Meissner effect and magnetic screening are observed in the respective field-cooled and zero-field-cooled traces (inset in Fig. 2). The critical current density ($J_c$) from field dependent magnetization (*M-H*) measurements at 6.5 K using the extended Bean Model [9] for $Sr_{0.57}Nd_{0.43}FeAsF$ was also studied. Fig. 2 shows the intragrain $J_c$ in this compound is in the range of $10^5$ A/cm$^2$, which is comparable to those of the iron oxypnictides.

* Corresponding author. Tel.: +81-29-8535297; e-mail: s_chong@ims.tsukuba.ac.jp.

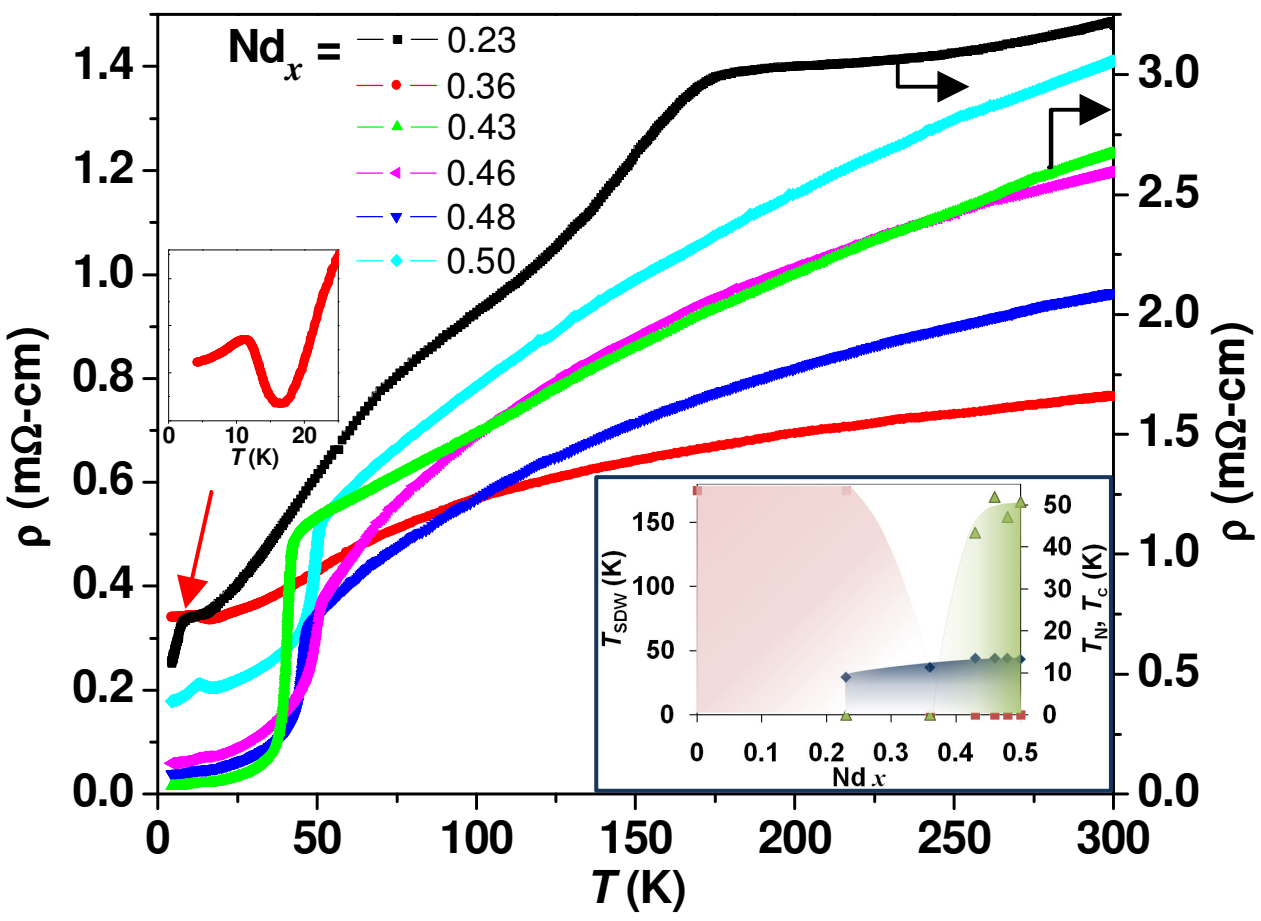

Fig. 1. *R-T* of $Sr_{1-x}Nd_xFeAsF$. The right inset shows a summary of the SDW transition ($T_{SDW}$,■), low temperature magnetic ordering ($T_N$,◆), and superconducting transition ($T_c$,▲) with respect to Nd-doping.

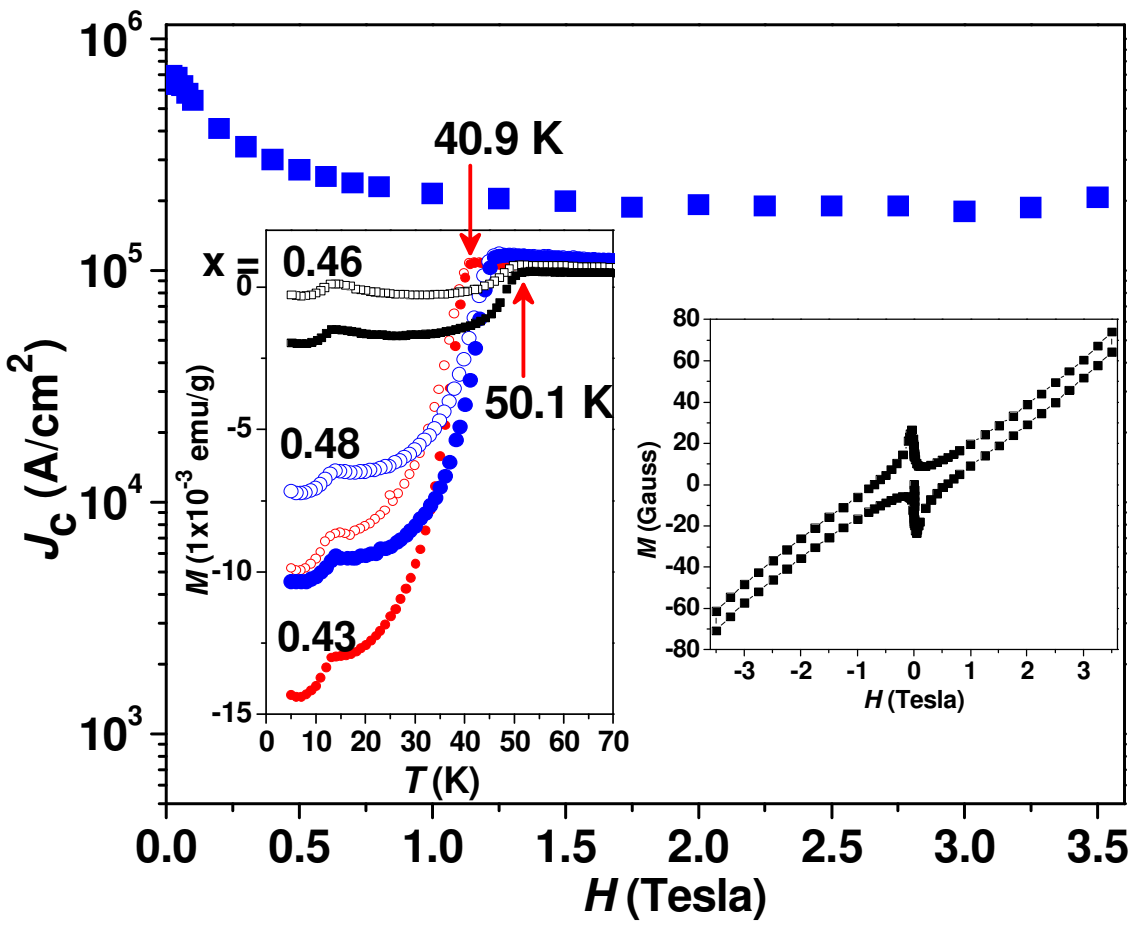

Fig. 2. *M-T* of superconducting $Sr_{1-x}Nd_xFeAsF$ samples at three different doping values (inset left; solid symbols – *ZFC*, open symbols – *FC*). The $J_c$ values (for $x$ = 0.43 at 6.5 K) are obtained from the inset *M-H* plot based on an average particle size of 15 μm.

In summary, superconductivity using Nd as a dopant in SrFeAsF with maximum $T_c$ close to 52 K has been achieved. All the superconducting samples show the complete suppression of the SDW anomaly at 175 K and the attenuation of the low temperature transition. Superconductivity is only realized above 0.4 Nd-doping which is in line with the current findings in other rare-earth-doped 1111-fluoroarsenides. $J_c$ in $Sr_{0.57}Nd_{0.43}FeAsF$ carried out at 6.5 K is in the range of $10^5$ A/cm$^2$, similar to the rare-earth oxypnictides.

## Acknowledgments

This work has been supported by JSPS-KAKENHI, Grant-in-Aid for Scientific Research (A) (18204031), the Ministry of Education, Culture, Sports, Science and Technology (MEXT), JAPAN, CREST JST, WPI at NIMS (MANA), JSPS Core-to-Core Program - Strategic Research Networks, "Nanoscience and Engineering in Superconductivity (NES)", and the JSPS postdoctoral fellowship for foreign researchers.